\documentclass[manuscript=article]{achemso}

\usepackage[version=3]{mhchem} % Formula subscripts using \ce{}
\usepackage{upgreek}
\usepackage{epsf}

\title{Microscopic Mechanism of Specific Peptide Adhesion to Semiconductor Substrates**
\newline\small{Michael Bachmann, Karsten Goede,* Annette G. Beck-Sickinger, Marius
Grundmann, Anders Irbäck, Wolfhard Janke}}

\author{}
\email{goede@physik.uni-leipzig.de}

\begin{document}

* Dr.~M.~Bachmann, Institut f\"ur Festk\"orperforschung, Theorie II,
Forschungszentrum J\"ulich, 52425 J\"ulich, Germany; and Institut f\"ur Theoretische Physik,
Universit\"at Leipzig, Postfach 100\,920, 04009 Leipzig, Germany; and Computational Biology \& Biological Physics, Department of Astronomy and Theoretical Physics, Lund University, S\"olvegatan 14A, 223 62 Lund, Sweden;

* Dr.~K.~Goede, Institut f\"ur Experimentelle Physik II, Universit\"at Leipzig, 
Linn\'estra{\ss}e 5,~04103 Leipzig, Germany, goede@physik.uni-leipzig.de, phone: +49 341 9732626, fax: +49 341 9732668;

* Prof.~A.~G.~Beck-Sickinger, Institut f\"ur Biochemie, Universit\"at Leipzig, Br\"uderstra{\ss}e 34, 04103 Leipzig, Germany;

* Prof.~M.~Grundmann, Institut f\"ur Experimentelle Physik II, Universit\"at Leipzig, 
Linn\'estra{\ss}e 5,~04103 Leipzig, Germany;

* Prof.~A.~Irb\"ack, Computational Biology \& Biological Physics, Department of Astronomy and Theoretical Physics, Lund University, S\"olvegatan 14A, 223 62 Lund, Sweden;

* Prof.~W.~Janke, Institut f\"ur Theoretische Physik, Universit\"at Leipzig, Postfach 100\,920, 04009 Leipzig, Germany

\vspace{0.5cm}

**We are thankful to Simon Mitternacht for helpful discussions regarding the peptide model and C. Dammann for peptide synthesis and purification. 
MB thanks
the DFG (German Science Foundation) and the Wenner-Gren Foundation (Sweden)
for research fellowships, and the German-Israel ``Umbrella'' program for support. 
MB, AI, and WJ are grateful for support by the
German-Swedish DAAD-STINT Personnel Exchange Programme. 
This work is also partially funded by the DFG under Grant  
Nos.\ JA 483/24-1/2/3, the Leipzig Graduate School of Excellence ``BuildMoNa'', TR 67 A4, and
the German-French DFH-UFA PhD College under Grant No.\ CDFA-02-07. 
Supercomputer time at the John von Neumann Institute for Computing (NIC), 
Forschungszentrum J\"ulich, is acknowledged (Grant Nos.\ hlz11 and jiff39).

%%%%%%%%%%%%%%%%%%%%%%%%%%%%%%%%%%%%%%%%%%%%%%%%%%%%%%%%%%%%%%%%%%%%%
%% The manuscript does not need to include \maketitle, which is
%% executed automatically.  The document should begin with an
%% abstract, if appropriate.  If one is given and should not be, the
%% contents will be gobbled.
%%%%%%%%%%%%%%%%%%%%%%%%%%%%%%%%%%%%%%%%%%%%%%%%%%%%%%%%%%%%%%%%%%%%%

\begin{abstract}
  The design of hybrid peptide--solid interfaces for 
nanotechnological applications such as biomolecular nanoarrays requires a deep understanding of the basic
mechanisms of peptide binding and assembly at solid substrates. Here we show
by means of experimental and computational analyses that the adsorption properties of 
mutated synthetic peptides at semiconductors exhibit a clear sequence-dependent
adhesion specificity. Our simulations of a novel hybrid peptide-substrate 
model reveal the correspondence between proline mutation and binding affinity to a clean silicon substrate. 
After synthesizing theoretically suggested amino-acid sequences with different binding
behavior, we confirm the relevance of the selective mutations upon adhesion
in our subsequent atomic force microscopy experiments. 
\end{abstract}


\section*{Introduction}

In the past few years, the interest in hybrid interfaces formed
by ``soft'' molecular matter and ``hard'' solid substrates has rapidly grown
as such systems promise to be relatively easily accessible candidates 
for novel biosensors or electronic devices. The enormous progress in
high-resolution microscopy and in biochemical
engineering of macromolecules
is the major prerequisite for studies of hybrid systems and 
potential applications\cite{sarikaya1,gray}. 
One particularly important problem 
is the self-assembly and adhesion of polymers, proteins, or protein-like 
synthetic peptides to solid materials such 
as, e.g., metals\cite{brown1,kremer1},
semiconductors\cite{whaley1,buriak1,goede1,goede2}, carbon and
carbon nano\-tubes\cite{hent2,wang}, and silica\cite{heinz2,broo1}.
Peptide and substrate specific binding affinity is particularly relevant in pattern recognition 
processes\cite{golumbfskie,bogner1}.
Systematic experimental studies have been performed to investigate binding properties 
of individual amino acids in their binding behavior to selected materials\cite{willett1}. 
Basic theoretical considerations of simplified polymer--substrate and protein--substrate models 
have predicted complex pseudophase diagrams\cite{bj1,bj3}. 

In bacteriophage display experiments, only a few peptides 
out of a library of $10^9$
investigated sequences with 12 amino acid residues were found to possess a particularly
strong propensity to adhere to (100) gallium-arsenide (GaAs) surfaces\cite{whaley1}. The
sequence-specificity of adsorption strength is a remarkable property, but the question remains
how it is related to the individual molecular structure of the peptides. We expect that
\emph{relevant} mutations of sites in the amino-acid sequence can cause a change of the binding
affinity. Indeed, one key aspect of our study is to show  
that proline is a potential candidate
for switching the adsorption propensities to cleaned (100) silicon (Si) substrates.

Silicon is one of the technologically most important semiconductors, as it serves, for example, 
as carrier substrate in microelectronics. For this reason, electronic 
and surface properties of Si are well investigated.
This regards, for example, oxidation processes in air\cite{frieser1,hemeryck1} and
water\cite{ling1,weldon1}, as well as the formation of hydride surface 
structures and the Si-binding characteristics
of small organic compounds\cite{buriak1,chabal1}.
\section*{Results and Discussion}
To guide the design of peptide--silicon interfaces, we first performed extensive
computer simulations of a novel hybrid model discussed below. 
For testing the theoretically revealed trends
of adsorption propensity changes by selected mutation, we have synthesized the suggested specific
mutants by means of multiple solid phase peptide synthesis. The theoretical predictions
were subsequently verified in
atomic-force microscope (AFM) experiments (see ~\ref{fig:fig1} and the detailed
descriptions in the Supplementary Material).

The hybrid model used in the computer simulations
is composed of two parts contributing to the energy $E({\bf X})$ of a peptide
conformation ${\bf X}$: the energy of the peptide as represented by the implicit-solvent all-atom
model introduced in Refs.\cite{irbaeck4,irbaeck5} and the 
interaction of the peptide with the substrate which is modeled in a coarse-grained 
way. The peptide model takes into account intrinsic
excluded volume repulsions between all atoms, a local potential 
which represents the interaction among neighboring NH and CO
partial charges, hydrogen bonding energy, and the interaction between
hydrophobic side chains\cite{irbaeck4,irbaeck5}.
The substrate model consists only of
atomic layers with surface specific atomic density and planar surface structure.
In this simplified model, 
each peptide atom feels the mean field of the atomic substrate layers.
The atomic density of these layers depends on the crystal orientation 
of the substrate at the surface.
Based on these assumptions, a generic noncovalent Lennard-Jones approach 
for modeling the interaction between
peptide atoms and surface layer is employed\cite{hent2,steele1}.
We have studied this model by means of multicanonical computer simulations\cite{muca1b} which provide us with 
canonical statistics for any temperature $T$. The partition function is thus given by 
$Z=\int {\cal D}{\bf X}\, e^{-E({\bf X})/RT}$,
where ${\cal D}{\bf X}$ is the formal functional integral measure for all possible conformations 
${\bf X}$ in the space
of the degrees of freedom. 
The statistical average of any quantity $O$ is 
$\langle O\rangle=Z^{-1}\int {\cal D}{\bf X}\, O({\bf X}) e^{-E({\bf X})/RT}.$
In our simulations, the integral is estimated by an average over a large set of conformations
(in each run about $10^9$ updates were performed) selected by multicanonical importance 
sampling. The precise modeling of the hybrid system and the multicanonical
simulation methodology are described
in the Supplementary Material.

The peptide with the amino acid sequence S1 [see ~\ref{fig:fig2}(a)] is a good example 
for the substrate specificity of adsorption. In recent comparative adsorption experiments,
it could be shown that although S1 binds strongly to GaAs(100), binding  
to Si(100) is very weak\cite{goede1,goede2}. In contrast, the adhesion
is strongly increased, if the Si substrate is oxidized. This can clearly be seen 
in ~\ref{fig:fig2}(a), where AFM images of S1 adhered at a de-oxidized (left)
and an oxidized Si(100) substrate (right) are shown. Peptide covered regions appear
bright. A quantitative measure for the binding propensity is the peptide adhesion 
coefficient (PAC), which is the relative area of the surface covered by peptide 
clusters\cite{goede1,goede2}. These PAC values are determined here by means of a cluster analysis of the respective AFM images. To reduce the dependence on the peptide concentration in solution, 
we here introduce the calibrated PAC (cPAC) as the ratio of PACs measured for the
binding of the peptides to Si(100) and GaAs(100) substrates under identical conditions.  
GaAs(100) is chosen as a reference substrate, since the peptides considered here bind comparatively
well to this substrate. The cPAC charts for S1 in ~\ref{fig:fig2}(b) clearly indicate the 
difference of binding affinity at cleaned and oxidized substrates. 

This is in contrast to sequence S3 [for sequence and AFM images see ~\ref{fig:fig2}(c)], 
which is a random permutation of S1 with unchanged
amino acid content. Surprisingly, the binding propensity of S3 to Si(100) was found to be
\emph{much} larger than that of S1\cite{goede1}. 
In this case, the binding affinities at cleaned and oxidized Si(100) substrates
are similarly strong, as the cPAC charts for S3 in ~\ref{fig:fig2}(b) show.
In recent computational analyses of the solvent properties
of these peptides, we have shown that also the folding behaviors in solution exhibit 
noticeable differences. This is also true at room temperature, where in both cases the population of 
the structurally different native folds is rather small\cite{msbji1}. 

Another remarkable result of this former computational study is that 
the qualitative folding behaviors of S1 and S3 are related to each other,
if these
sequences are pairwisely mutated at the position of proline, which occurs once in the sequences S1 and 
S3\cite{msbji1}. The mutated sequence S1' differs from S1 only by the exchange 
of proline (P) at position 4 and threonine (T)
at 9. 
Similarly, in S3', proline at 9 is exchanged with the aspartic acid (D) at 4, 
compared to S3 [see ~\ref{fig:fig3}(a)].
These replacements were reasoned by the guess that the particular steric properties of proline,
and thus its place in the sequence, influence the folding. It turned out that 
the folding behavior of S1' in solution is close to that 
of S3, whereas S3' behaves rather like S1\cite{msbji1}.
Before we can address the question whether these results 
are also of essence for the adsorption behavior to Si(100), we need to discuss microscopic 
properties of de-oxidized Si(100) substrates.

In our experiments, the Si(100) surfaces were first cleaned in 
a solution of ammonium fluoride (NH$_4$F) and hydrofluoric acid 
(HF)\cite{goede1,goede2}. Then, the peptide adsorption process
took place in
de-ionized water. This standard procedure ensures that the Si surface is virtually
free of oxide and possesses strongly hydrophobic properties\cite{frieser1,ling1}
(for sample preparation and experimental details see the Supplementary Material). 
The initial
Si--F bonds after etching are replaced by Si--H bonds in the rinsing process in
de-ionized water. After drying the sample, AFM scans of the surface were performed. 
Although the oxidation also proceeds in water\cite{ling1,weldon1},
there are clear indications (maximum water droplet contact angle after removing
the samples off the peptide solution) that the hydrophobicity
of the Si samples remains largely intact during the peptide adsorption process.
It is also known that Si surfaces are comparatively rough after HF treatment\cite{chabal1}.
Thus, the  
reactivity of the surface is influenced by steps, which depend on the offcut and its directions.
This renders an atomistic modeling intricate, even more as 
Si(100)-2$\times$1 surfaces are also known to form Si--Si dimers
on top of the surface\cite{buriak1} with highly reactive dangling bonds. From the considerations
and the experimental preparations described above, it seems plausible that these bonds are 
mainly passivated by hydrogen, forming hydride layers\cite{buriak1,ling1,chabal1}. 
It should be emphasized that under these conditions the surface structure of Si(100) is
substantially different from oxidized Si(100) which is polar and in effect 
hydrophilic\cite{frieser1}. 
An important result of ~\ref{fig:fig2} is that the binding of S1 and S3 to \emph{oxidized} 
GaAs(100) and Si(100) surfaces is virtually
independent of the substrate type (cPAC $\approx 1$). Thus, the top oxygen layer screens the
substrate from the peptide. 
The different adhesion propensities to the \emph{clean} (hydrated) substrates [see also ~\ref{fig:fig2}] 
lead to the 
conclusion that oxidation has not yet strongly progressed during the peptide adsorption
process. We 
conclude that the key role of water is the slowing down of the oxidation
process of the Si(100) surface, but for the actual binding process its influence is rather
small. In particular, we do not expect that stable water layers 
form between adsorbate and substrate.    

These characteristic properties of HF treated Si(100) surfaces in de-ionized water effectively
enter into the definition of our hybrid model of the peptide--silicon interface
(for details see the ``Model and Methods'' section in the Supplementary Materials)
which serves as the basis for our theoretical analysis 
and interpretation of the specificity of peptide adhesion on these interfaces.

In order to quantify the degree of adsorption, we define the ratio of heavy (non-hydrogen)
atoms located in a distance $z_i\le 5\:\textrm{\AA}$ from the substrate, $n_h$, and the total number
of heavy atoms, $N_h$, as the adsorption parameter $q=n_h/N_h$. The temperature dependence of
its relative change by proline mutation, $\Delta q({\rm S}n\to{\rm S}n')=
(\langle q({\rm S}n')\rangle-\langle q({\rm S}n)\rangle)/\langle q({\rm S}n)\rangle$ (with $n=1,3$), 
is shown in ~\ref{fig:fig3}(b). The main result is that due to
this selective mutation, the Si(100) adsorption affinity from S1 to S1' increases 
[$\Delta q({\rm S1}\to{\rm S1'})\approx 0.11$ at $T=300$ K], 
while it decreases by about the same amount as S3 is mutated to S3' 
[$\Delta q({\rm S3}\to{\rm S3'})\approx -0.15$ at $T=300$ K]. This result is directly connected with
the tendency to form secondary structures.  
In ~\ref{fig:fig3}(c), the respective $\alpha$-helix content
[ratio of the dihedral Ramachandran
angles of the inner 10 residues satisfying $\phi\in(-90^\circ,-30^\circ)$ and
$\psi\in(-77^\circ,-17^\circ)$]
and $\beta$-strand content 
[dihedral angles in the intervals $\phi\in(-150^\circ,-90^\circ)$ 
and $\psi\in(90^\circ,150^\circ)$] of the \emph{bound}
conformations are shown. We define a conformation as bound to the substrate, if
at least 2\% of the heavy atoms are within a 5 \AA\/ distance from the surface. 
There is a clear tendency that residues of S1 and S3' are rather in $\alpha$ state
and residues of S3 and S1' in $\beta$ state. Yet the small secondary-structure contents
are quite similar to what we found
for the peptides in solution (without substrate)\cite{msbji1}, which were qualitatively 
consistent
with analyses of CD spectra\cite{goede2}. It is a noticeable result that here 
secondary structures are not stabilized near the cleaned Si(100) substrate, whereas in recent 
adsorption experiments of a synthetic peptide binding at silica nanoparticles, 
a stabilization of $\alpha$-helices was observed~\cite{broo1}. 

The experimental results shown in ~\ref{fig:fig3}(d) confirm that 
the proline mutation of S1 indeed increases the Si(100) binding affinity, while 
an analogous mutation
decreases the binding strength of S3 by about the same value. The AFM images
indicate the increased substrate coverage for S1' and the decrease in the case of S3' [cp.\
with the corresponding images in ~\ref{fig:fig2}(a) and (c), respectively]. By measuring
the associated cPAC values, we find that $\Delta {\rm cPAC}({\rm S1}\to {\rm S1'})=
{\rm cPAC}({\rm S1'})-{\rm cPAC}({\rm S1})\approx 0.27$
and $\Delta {\rm cPAC}({\rm S3}\to {\rm S3'})\approx -0.25$. This convincingly confirms our 
theoretical prediction from
the hybrid-model simulations.
\section*{Conclusions}
In summary, we have predicted by computer simulations and verified by AFM experiments that a selected
proline mutation of short peptides facing a de-oxidized silicon substrate can substantially
change the binding affinity in a very predictive and specific way. We could also show that 
this behavior is in part due to a qualitatively different folding behavior of the
mutated sequences in the vicinity of the substrate. The proline position most likely also affects the aggregation properties\cite{goede2} of the peptides and thereby indirectly again their binding characteristics. Building up on simulations of single-molecule behavior, as those discussed in the present manuscript, simulating coupled folding and aggregation while binding will therefore constitute a rewarding future project.
Gaining deeper insights into the general principles of binding specificities 
is a first fundamental step towards the design of nanosensors with specific biomedical
applications. Thus, the extension of our study to biomolecules is natural and the
identification of unique bioprotein adsorption signals in experiments with nanoarrays
of several materials is a prerequisite for future applicability of such hybrid systems in biotechnology.
%



\vspace{0.5cm}
The precise modeling of the hybrid system, the multicanonical
simulation methodology and details of the peptide selection, the AFM experiments and the sample preparation are described
in the Supplementary Material.

%\bibliography{achemso}
%\textsf{natbib}

\newpage

\begin{figure*}
\centerline{\epsfxsize=8.8cm \epsfbox{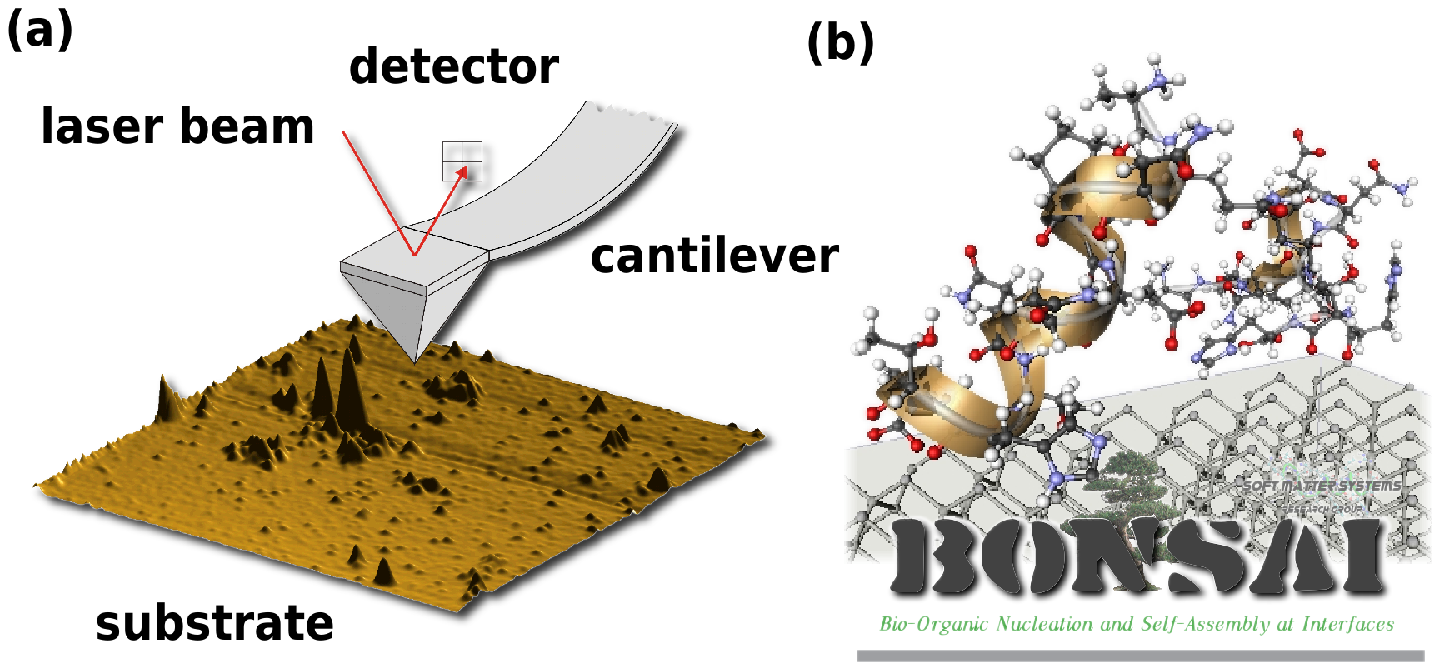}}
\caption{\label{fig:fig1} 
\textbf{Methodologies.} (a) Sketch of the atomic-force microscope. The original AFM image exhibits
S1 peptide clusters on an oxidized 10$\times$10 $\upmu$m$^2$ silicon substrate. 
The height of the highest cluster
is 56 nm. (b) Computer simulations were performed with the BONSAI 
package that we developed for Monte Carlo simulations of peptide models. 
The snapshot shows S1 peptides forming helical segments near a silicon substrate. 
}
\end{figure*}

\newpage

\begin{figure*}[!htb]
\centerline{\epsfxsize=8.8cm \epsfbox{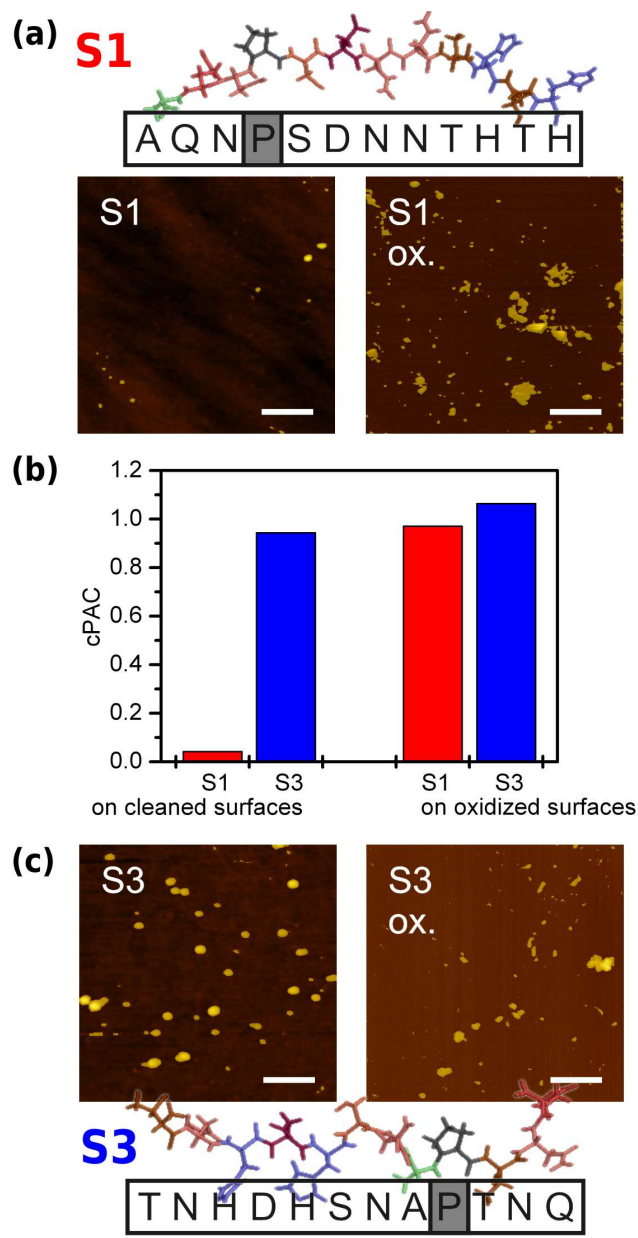}}
\caption{\label{fig:fig2} 
\textbf{Adsorption to cleaned and oxidized substrates.} Exemplified AFM images of peptides (a) S1 and (c) S3
adsorbed to cleaned (de-oxidized) 
and oxidized Si(100) surfaces. The AFM scale bar is 1 $\upmu$m.
(b) Calibrated peptide adhesion coefficients (cPAC)
for S1 and S3 adsorption to cleaned and oxidized Si(100) substrates. Single-letter code of amino 
acids occurring in the peptides of our study: A -- alanine, D -- aspartic acid, H -- histidine,
N -- asparagine, P -- proline, Q -- glutamine, S -- serine, T -- threonine.
}
\end{figure*}
\newpage

\begin{figure*}[!htb]
\centerline{\epsfxsize=15cm \epsfbox{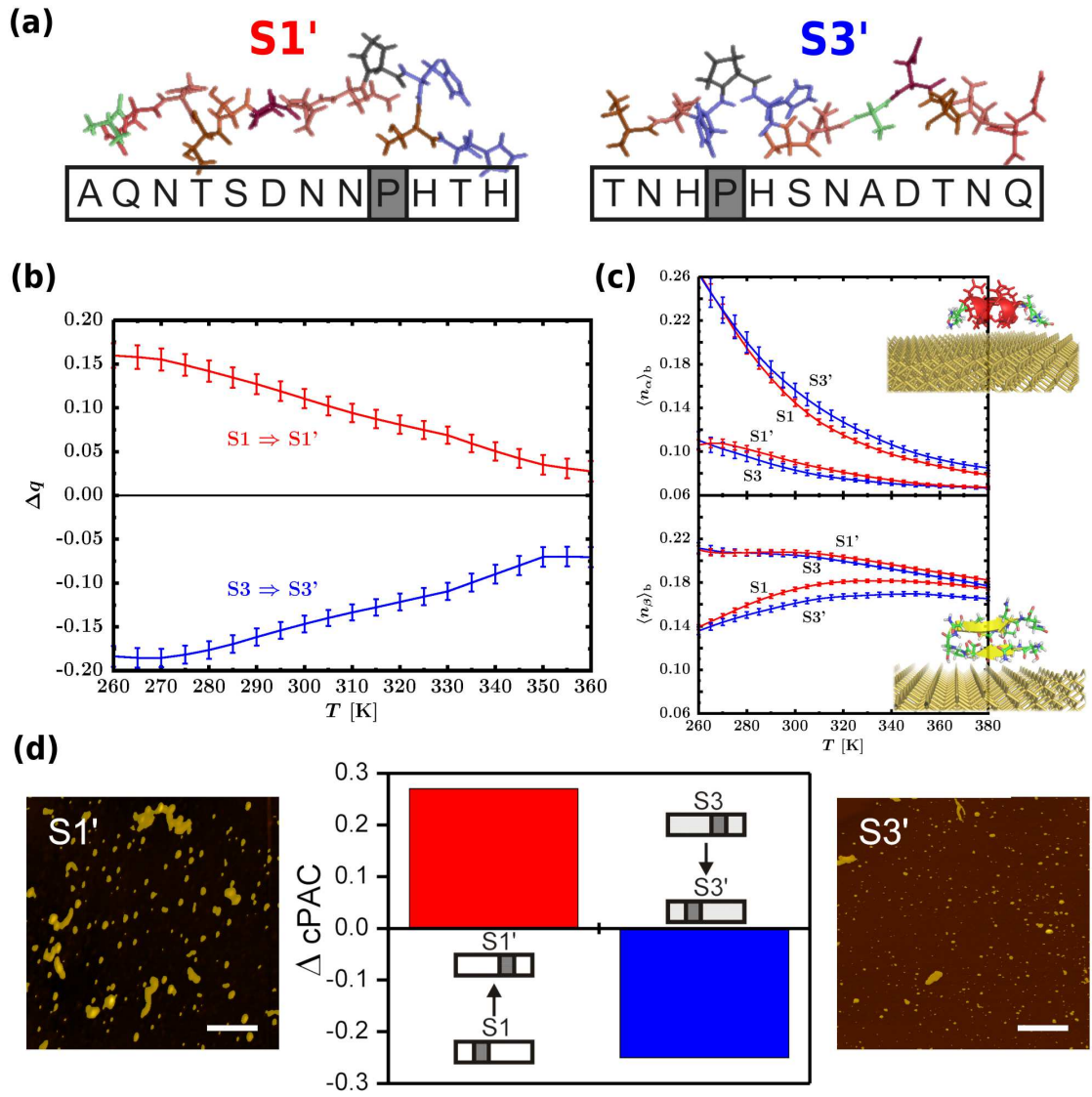}}
\caption{\label{fig:fig3} 
\textbf{Reversed adsorption propensity of proline-mutated peptides.} 
(a) Proline-mutated sequences S1' and S3'. (b) Adsorption parameter $\Delta q$ as a function of temperature
from our computer simulations. The prediction is that the proline mutation of S1 causes an increase 
of binding affinity, whereas mutating S3 leads to a decrease, i.e., proline mutation reverses the binding 
propensity of S1 and S3. (c) $\alpha$-helix content $\langle n_\alpha \rangle_{\rm b}$ and 
$\beta$-strand content $\langle n_\beta \rangle_{\rm b}$ of \emph{bound} conformations
also exhibit a pairwise reversal of the tendency to form secondary structures. Exemplified 
conformations depicted in the insets are lowest-energy structures identified in the simulations
(with rather small populations at
room temperature) and represent the preferred trends in secondary-structure formation: 
helical for S1 and S3', sheet-like for S1' and S3. (d) Confirmation by AFM experiments at 
room temperature. 
The cPAC increases by proline-mutating S1 and decreases by
about the same value if S3 is mutated. AFM scale bar 1 $\upmu$m.
}
\end{figure*}

\newpage

\begin{figure*}[!htb]
\center{\includegraphics[width=1.0\textwidth]{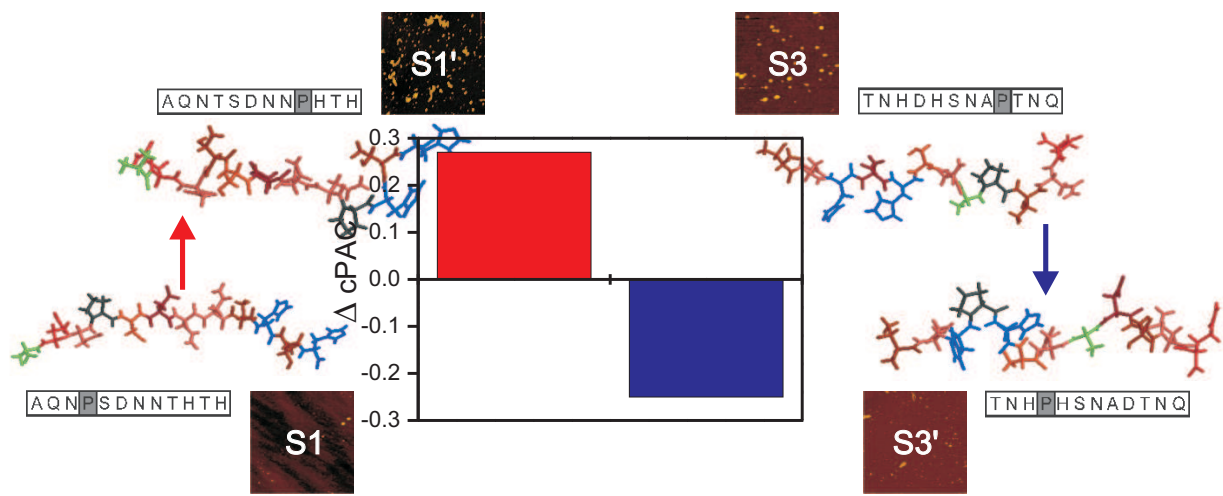}}
\caption{\textbf{Suggested TOC graphic.} Designing novel hybrid peptide-solid interfaces for nanotechnology requires a sound understanding of peptide
binding and assembly at substrates. Here we show by means of experimental and computational
analyses for a novel hybrid peptide-substrate system that the adsorption properties of mutated synthetic peptides at semiconductors
exhibit a clear sequence-dependent adhesion specificity, which is mainly governed by the positions of crucial amino acids.}
\label{fig:TOC}
\end{figure*}

\end{document}